\documentclass[twocolumn,superscriptaddress,showpacs,prd]{revtex4-1}
\usepackage{epsfig,graphicx,color,appendix,amssymb,fge}
\usepackage{amsmath}
\usepackage{mathtools}

\begin{document}

\title{\mbox{}\\[1pt]
 Towards a Model of Quarks and Leptons}

\author{Y. H. Ahn}
\affiliation{Institute for Convergence Fundamental Study, Seoul National University of Science and Technology, \\
       Seoul 139-743, Korea}

\author{Sin Kyu Kang}
\affiliation{School of Liberal Arts, Seoul National University of Science and Technology,
       Seoul 139-743, Korea}

\author{Hyun Min Lee}
\affiliation{Department of Physics, Chung-Ang University, Seoul 06974, Korea.}


\begin{abstract}
\noindent We propose an extra-dimension framework on the orbifold $S^1/Z_2$ for understanding the origin of the fermion mass and mixing hierarchies.  Introducing the flavor symmetry $G_F(=${\it non-Abelian}$\times${\it Abelian}) as well as the extra gauged $U(1)$ symmetries through the bulk, we regard the $SU(2)$ singlet and doublet fermions in the Standard Model (SM) to be localized at the separate 3-branes and let the extra $SU(2)$ singlet flavored fermions in the bulk couple to the SM fermions at the 3-branes.
The extra $U(1)$ symmetries satisfy the $U(1)$ gravitational anomaly-free condition, playing a crucial role in achieving the desirable fermion mass and mixing hierarchies and making the flavored axion naturally light.
 The singlet scalar fields, the flavon fields, are responsible for the spontaneous breaking of $G_F$ on the two 3-branes, while the $SU(2)$ singlet flavored fermions are integrated out to give rise to the effective Yukawa couplings for the SM fermions, endowed with the information of $G_F$ breaking in the two sectors. The flavored axion from the PQ symmetry is also proposed for solving the strong CP problem and being a dark matter candidate in our model.
\end{abstract}
\maketitle %
%%%%%%%%%%%%%%%%%%%%%%%%%%%%%%%%%%%%%%%%%%%%%%%%%%%%%%%%%%%%%%%%%%%%%%%%%%%%%%%%%%%%%%%%%%
\section{Introduction}
The observed hierarchies in the masses and mixings of quarks and leptons are one of the most puzzling problems in particle physics.
A plausible explanation is to introduce a new gauge symmetry which  is spontaneously broken at some ultraviolet (UV) scale and leaves behind its global subgroup. In this case, in the low-energy effective theory, the symmetry structure is composed of the Standard Model (SM) gauge symmetry $G_{\rm SM}=SU(3)_C\times SU(2)_L\times U(1)_Y$ and a remnant global symmetry which plays an essential role in making desirable flavor structure of quarks and leptons \cite{Ahn:2016typ}.
It is known that flavor-dependent $U(1)$ gauge symmetries and/or non-Abelian discrete symmetries can arise from the isometry of the compactified extra dimensions in string theory. In addition,  the compactification of the extra-dimensions can be accompanied by certain 3-branes (4 dimensional (4D) surfaces embedded in higher dimensional spaces)\cite{Horava:1995qa}. 

Inspired by some string compactifications,
in this paper, we propose a mechanism for generating the fermion mass and mixing hierarchies in the  {\it extra-dimension} framework
with the flavor symmetry.
The extra gauged $U(1)$ symmetries are also introduced under the $U(1)$ gravitation anomaly-free condition and they are crucial to achieve the desirable fermion mass and mixing hierarchies in this scenario.
These are the features in this work, which are distinguishable from other approaches to tackle the flavor problem in the extra-dimension framework
\cite{Mirabelli:1999ks,Gherghetta:2000qt,Kaplan:2000av,Huber:2000ie,Agashe:2004cp,Casagrande:2008hr,Agashe:2008fe,Archer:2012qa,Frank:2015sua,Ahmed:2019zxm,Wojcik:2019kfq,Kang:2020cxo}. A simple toy model with vector-like leptons for seesaw leptons has been recently proposed in light of the muon $g-2$ anomaly \cite{Lee:2021gnw}, but without involving a flavor symmetry.

For our purpose, we construct a higher dimensional theory compactified on the orbifold $S^1/Z_2$ with a global symmetry group  for flavors, $G_{F}=U(1)\times${\it non-Abelian finite group}, which might be originated from certain string compactifications.
A set of SM gauge singlet scalar fields ${\cal F}$, charged under $G_F$, the so-called {\it flavon fields}, are  located at two 3-branes in the extra dimension.
In the 4D effective Lagrangian, the flavor fields act on dimension-four(three) operators well-sewed by $G_F\times G_{\rm SM}$ at different orders to generate the effective interactions for the SM and the right-handed neutrinos, as follows;
\begin{eqnarray}
 &{\cal L}_{4D}=\tilde{c}_1{\cal O}_3\,{\cal F}\,\tilde{{\cal X}}+{\cal O}_4\sum^{\rm finite}_{n=0}c_n\Big(\frac{\cal F}{M_5}\Big)^n\,{\cal X}_n\,,
\label{4dL}
\end{eqnarray}
where ${\cal O}_{3(4)}$ are dimension-3(4) operators, $\tilde{c}_1(c_n$) are the complex coefficients of order unity, and $\tilde{{\cal X}}({\cal X}_n$) are dimensionless parameters induced due to the non-local effects of bulk messenger fields.  
${\cal F}$ acquire vacuum expectation values (VEVs) $\langle{\cal F}\rangle$ from some dynamics, thereby breaking $G_F$.
Then, the vacuum structure of  the flavons  plays a crucial role in achieving  the SM fermion mass and mixing hierarchies.

As a bonus in our scenario, the pseudo-Goldstone modes coming from the flavon fields are localized on the 3-branes, becoming candidates for flavored-axions $A_i$ (and a QCD axion) \cite{Ahn:2014gva} with decay constants determined by $\langle{\cal F}\rangle$.
However, it is well known that non-perturbative quantum gravitational anomaly effects \cite{Kamionkowski:1992mf} could spoil the axion solution to the strong CP problem.
In order to keep the axion solution in our scenario, we  need to suppress the explicit breaking effects of the axionic shift symmetry by gravity and consistently couple gravity to matter. To this, we impose the $U(1)$-mixed gravitational anomaly-free conditions for the extra gauged $U(1)$ symmetries, in turn, obtaining the constraints on the $U(1)$ charges of quarks and leptons.

\section{Flavor physics embedded into 5D theory}
We consider a 5D theory for flavors compactified on the orbifold $S^1/Z_2$ where the extra-dimension on the circle $S^1$ is identified $y$  with $-y$ \cite{Randall:1999ee}. The orbifold fixed points at $y=0$ and $y=L$, the boundaries of the 5D spacetime, are the locations of two 3-branes, We assume that all  the ordinary matter fields localized at either brane  and they are charged under the flavor symmetry $G_F$.
We specify $G_F=SL_2(F_3)\times U(1)_X$ where $SL_2(F_3)$ is the symmetry group of the double tetrahedron\cite{Feruglio:2007uu}.

The metric solution to the 5D Einstein's equations, respecting the 4D Poincare invariance in the $x^\mu$ direction, is given by
\begin{eqnarray}
ds^2=e^{2\sigma(y)}\eta_{\mu\nu}dx^{\mu}dx^{\nu}-dy^2
\label{met1}
\end{eqnarray}
where the extra dimension is compactified on an interval $y\in[0,L]$, 
the warp factor is given by $\sigma(y)=ky$ with $k=\sqrt{-\frac{\Lambda}{6M^3_5}}>0$ and $\Lambda$  being the bulk cosmological constant, and the 4D Minkowski flat metric is $\eta_{\mu\nu}={\rm diag}(+,-,-,-)$. Note that, in Eq.(\ref{met1}), we can always take $\sigma(0)=0$ by rescaling the $x^\mu$ coordinates. Then, the 4D reduced Planck mass $M_P\simeq2.43\times10^{18}$ GeV can be extracted in terms of the 5D Planck mass $M_5$ as
\begin{eqnarray}
&M^2_P=M^3_5\int^{L}_{-L}dy\,e^{2\sigma(y)}=\frac{M^3_5}{k}\big(e^{2kL}-1\big)\,
\label{re_M}
\end{eqnarray}
where $M_5$ is assumed to be higher than the electroweak scale,  but $M_5$ and $k$ are lower than $M_P$. Then, the scale of flavor dynamics would be given by the UV cutoff $M_5$.

For exchanging the information of the breaking of flavor symmetry $G_F$ between the two 3-branes, we introduce 
$SU(2)$ singlet {\it flavored bulk fermions} $\Psi_{f_i}$ and
their mirror partners $F_{f_i}$ charged under $G_{\rm SM}\times G_F$, propagating in a 5D space with the metric Eq.(\ref{met1}).
Here, $f_i=u_i$ (up-type quark), $d_i$(down-type quark), $l_i$ (charged lepton), and $F_{f_i}$ is introduced to keep the conservation of electric charge.
While $\Psi_{f_i}$ with hypercharges $Y_f$ and masses $M_{f_{i}}$ interact with the normal matter fields confined at $y=0$ or $L$ brane,
$F_{f_i}$ with hypercharges $-Y_f$ and masses $M_{f_{i}}$ does not do so due to $U(1)_Y$ symmetry.
Unlike Ref.\cite{Goldberger:1999uk}, in this work, the compactification length $L$ can be constrained thanks to the introduction of flavored bulk fermions. In Table-\ref{reps_v}, for  flavon fields ${\cal F} =\Phi_S, \Theta (\tilde{\Theta}), \Phi_T, \rho, \eta, \chi, \tilde{\chi}$ and flavored bulk fermions $\Psi_f$, we present the representations of $SL_2(F_3)$ and quantum charges under $ U(1)_{X_1}\times U(1)_{X_2}\times U(1)_{X_T}$.

%\begin{widetext}
\begin{table}[t]
%\begin{center}
\caption{\label{reps_v} Representations and quantum charges of SM singlet flavon fields and bulk fermions under $G_F$.}
%\begin{ruledtabular}
\begin{tabular}{cccccc}
Field\ &$SL_2(F_3)$&$U(1)_{X_1}$&$U(1)_{X_2}$ & $U(1)_{X_T}$  & brane(y) \\
\hline
$\Phi_S$ : $\Theta(\tilde{\Theta})$&${\bf 3}$: ${\bf 1}$  & $1$ & $0$ & $0$ & $0$ \\
%$\Theta(\tilde{\Theta})$ & ${\bf 1}$ & $1$ & $0$ & $0$\\
$\Phi_T$ : $\varrho$ & ${\bf 3}$ :  ${\bf 1}$ & $0$&$0$ & $1$, $-1$& $L$\\
%$\varrho$& ${\bf 1}$ & $0$ & $0$ & $-1$\\
$\eta$& ${\bf 2}'$ & $0$ & $0$ & $0$& $L$\\
$\chi$, $\tilde{\chi}$& ${\bf 1}$ & $0$ & $1, -1$ & $0$& $L$ \\
\hline
$\Psi_l$ & ${\bf 3}$ & $-13/2$ & 0 & 0 &  \\
$\Psi_{u_1}:\Psi_{d_1}$ & ${\bf 1}$ & $-6$ : $-5$ & 0 & 0 &  \\ 
$\Psi_{u_2}:\Psi_{d_2}$ & ${\bf 1^{\prime}}$ & $-6$ : $-5$ & 0 & 0 &  \\ 
$\Psi_{u_3}:\Psi_{d_3}$ & ${\bf 1^{\prime \prime}}$ & 0 : 3 & 0 & 0 &  \\ 
\end{tabular}
%\end{ruledtabular}
%\end{center}
\end{table} 
%\end{widetext}
%
All ordinary matter and Higgs fields charged under $G_{\rm SM}$ with +1 and 0 charges under $U(1)_R$, respectively, are localized at either brane.
Then, all the SM fermion mixings and masses can be generated by non-local effects involving both branes and local breaking effects of $G_F$  due to flavon fields.
For the orbifold compactification, we set all elementary fermions form a chiral set, and their group representations and quantum numbers are summarized in Table-\ref{reps_f}.
From  the $U(1)_{X_k}\times[SU(3)_C]^2$ anomaly coefficient defined by $\delta^{\rm G}_k\delta^{ab}=2\sum_{\psi_f}X_{k\psi_f}{\rm Tr}(T^aT^b)$
in the QCD instanton backgrounds with $T^a$ and $X_k$ being $SU(3)_C$ generators and $U(1)_{X_k}$ charge, respectively, we get
$\delta^{\rm G}_1=-9$ and  $\delta^{\rm G}_2=-11$ with the domain-wall number $N_{\rm DW}=1$.
In this model, $U(1)_{X_1}\times U(1)_{X_2}\equiv U(1)_{\tilde{X}}$ is a pure axial symmetry  $U(1)_{\rm PQ}$.
A color anomaly coefficient $N_C$ of $U(1)_{\tilde{X}}\times[SU(3)_C]^2$  and  an electromagnetic one $E_A$ of $U(1)_{\tilde{X}}\times[U(1)_{\rm EM}]^2$  are defined by  $N_C\equiv2{\rm Tr}[\tilde{X}_{\psi_f}T^aT^a]=2\delta^{\rm G}_1\delta^{\rm G}_2$ and $E_A=2\sum_{\psi_f} \tilde{X}_{\psi_f}(Q^{\rm em}_{\psi_f})^2$ with $Q^{\rm em}_{\psi_f}$ being the $U(1)_{\rm EM}$ charge of field $\psi_f$, respectively.
Then, their ratio becomes $E_A/N_{C}=-761/99$. 
\begin{table}[h]
%\begin{center}
\caption{\label{reps_f} Representations of quark, lepton, and electroweak two Higgs $H_{u(d)}$ fields under $SL_2(F_3)\times U(1)_{X_i}$ ($i=1,2,T$). All fields are left-handed particles/antiparticles. All of them have zero $U(1)_R$. }
%\begin{ruledtabular}
\begin{tabular}{cccccc}
Field &$SL_2(F_3)$&$U(1)_{X_1}$&$U(1)_{X_2}$ & $U(1)_{X_T}$ & brane   \\
\hline
$Q_1$, $Q_2$, $Q_3$&${\bf 1}$ ${\bf 1}'$ ${\bf 1}''$& $-8,-6, 0$ & $0,0,0$ & $0,0,0$ & $y=0$ \\
${\cal D}^c$, $b^c$&${\bf 2}'$ ${\bf 1}'$&$5$, $-3$&$-14$, $18$ & $0,0$ & $y=L$\\
${\cal U}^c$, $t^c$&${\bf 2}'$ ${\bf 1}'$&$6$, $0$&$-6$, $11$ & $0,0$ & $y=L$\\
$L$&${\bf 3}$&$-\frac{15}{2}$ & $0$ & $0$ & $y=0$\\
$e^c$, $\mu^c$, $\tau^c$ & ${\bf 1}$, ${\bf 1}''$, ${\bf 1}'$ & $\frac{13}{2},\frac{13}{2},\frac{13}{2}$ & $41$, $31$, $21$ & $-2$, $1$, $1$ & $y=L$\\
$N^c$&${\bf 3}$&$-\frac{1}{2}$ & $0$ & $0$ & $y=0$\\
\hline
$H_{u(d)}$& ${\bf 1}$ & $0$ & $0$ & $0$ & $y=0$\\
\end{tabular}
%\end{ruledtabular}
%\end{center}
\end{table} 

The flavor symmetry $G_F$ is spontaneously broken by the nontrivial  VEVs of flavons.
Note that the $U(1)$ charges of the fields are determined so as to satisfy the $U(1)$ gravitation anomaly-free condition and the empirical hierarchies of fermion masses and mixings.
 For instance, in a supersymmetric model, the brane-localized superpotential for flavons with $G_F$ invariance is given in Eq.(\ref{d_pot}).
From the minimization conditions of the $F$-term scalar potentials, the VEVs of $\Phi_S$ and $\tilde{\Theta}$ localized at $y=0$ brane are obtained
{\small\begin{eqnarray}
  \langle\Phi_S\rangle=\frac{v_S}{\sqrt{2}}(1,1,1)\,,\quad  \langle\Theta\rangle=\frac{v_\Theta}{\sqrt{2}}\,, \quad  \langle\tilde{\Theta}\rangle=0\,,
  \label{vev1}
\end{eqnarray}}
with $\kappa=v_S/v_\Theta$ in supersymmetric limit. 
For $\Phi_T$, $\varrho$, $\chi(\tilde{\chi})$, and $\eta$ localized at $y=L$ brane
{\small\begin{eqnarray}
    &&\langle\Phi_T\rangle=\frac{v_T}{\sqrt{2}}(1,0,0)\,,\qquad\qquad \langle\varrho\rangle=\frac{v_\varrho}{\sqrt{2}}\,,\nonumber\\
    &&~~\langle\chi\rangle=\langle\tilde{\chi}\rangle=\frac{v_\chi}{\sqrt{2}}\,,\qquad\quad~ \langle\eta\rangle=\frac{v_\eta}{\sqrt{2}}(1,0)\,.
  \label{vev2}
\end{eqnarray}}
Denoting $\nabla_{\cal F}\equiv v_{\cal F}/(\sqrt{2}M_5)$ and following the procedure in Ref.\cite{Ahn:2018cau}, we obtain  
\begin{eqnarray}
\nabla_\Theta=\nabla_S/\kappa=\big|\delta^{\rm G}_1/\delta^{\rm G}_2\big|\sqrt{2/(1+\kappa^2)}\nabla_\chi.
\label{ab}
\end{eqnarray}
The brane-localized Yukawa Lagrangian for up- and down-type quarks and leptons, which is invariant under $G_{\rm SM}\times G_F$, involve  the exchanges of their flavored bulk fermions, but no tree-level interactions in the SM:
\begin{widetext}
\begin{eqnarray}
&{\cal L}_Y=\delta(y)\big\{\sum\limits_{i}^{3} Z^u_{ii}\Psi^c_{ui}Q_iH_u+Z^d_{ii}\Psi^c_{di}Q_iH_d
+\sum\limits_{ i>j}^{3} Z^d_{ij}\Psi^c_{di}Q_j(\Phi_S\Phi_S)_{j}\frac{H_d}{M^2_5}+\big(y_{\nu}\frac{\Theta}{M_5}+\tilde{y}_{\nu}\frac{\Phi_S}{M_5}\big)  N^cLH_u\nonumber\\
&+Z_{L}(\Psi^c_{\ell}L)_{{\bf1}}H_d+
\frac{1}{2}(y_R \Phi_S +y_\Theta\Theta+y_{\tilde{\Theta}}\tilde{\Theta})(N^cN^c)\big\}+\delta(y-L)\big\{\hat{Z}_{t3}t^c\Psi_{u3}+Z_{u2}{\cal U}^c\eta\Psi_{u2}\nonumber\\
&+Z_{u1}{\cal U}^c\eta\frac{\Phi_T\varrho}{M^2_5}\Psi_{u1}
+(t\rightarrow b, {\cal U}\rightarrow{\cal D}, u\rightarrow d)
+Z_{e}e^c\Psi_{\ell}\frac{\Phi_T\Phi_T}{M_5}+\sum_{\omega=\mu,\tau}Z_{\omega}\omega^c\Psi_\ell\Phi_T\frac{\varrho^2}{M^2_5} \big\}\,,
\label{Yuka1}
\end{eqnarray}
\end{widetext}
%\end{widetext}
where
higher order terms combined by ($\eta\eta$, $\Phi_T\Phi_T$, $\Phi_T\varrho$) are omitted.
In Eq.(\ref{Yuka1}),   $\{Z^f_{ij}, Z_L, Z_{fi}, Z_{e,\mu,\tau}\}$ and $\hat{Z}_{fi}$ are functions of flavons, {\it i.e.} $Z= Z[(\frac{\cal F}{M_5})^p]$ with $p$ being a combination of $U(1)_{X_i}$ charges assigned so as for ${\cal L}_Y$ invariant under $U(1)_{X_i}$, and their mass dimensions
are $-1/2$ and $+1/2$, respectively.
As will be shown later, the exponent $p$ is responsible for the hierarchies of fermion masses and mixings.
The coupling  $y_{\nu} (\tilde{y}_\nu$) and $y_{R, (\Theta,\tilde{\Theta})}$ are respectively a function of ${\cal F}$ and their mass dimensions are all zero.

\section{4D effective theory}
For the effective theory in 4D, we expand the left- and right-handed bulk fermions  in terms of  Kaluza-Klein(KK)  modes as 
%\begin{eqnarray}
$\Psi_{f_i L(R)}(x, y)=\frac{e^{-\frac{3}{2}\sigma(y)}}{\sqrt{L}}\sum\limits_{n}\psi^{n}_{f_iL(R)}(x)\,f^n_{iL(R)}(y)$.
%\label{KKw1}
%\end{eqnarray} 
They are chosen to obey the equation of motion coming from the 4D action $S=\sum\limits_{n} \int d^4x\bar{\psi}^{n}_i(f_i\gamma_\mu D^\mu-m_f^n)\psi^{n}_{f_i}$ where $m^n_f$ is the 4D mass of the $n$-th KK mode, with the normalization condition
$\frac{1}{L}\int^L_0dy\,f^m_{iL(R)}f^n_{iL(R)}=\delta_{mn}$.
We choose a gauge $A_5=0$
such that the KK modes are independent of the gauge fields. In terms of left- and right-handed spinors $f^n_{iL,R}$, we obtain
\begin{eqnarray}
&\big(\partial_y+\frac{1}{2}\sigma'-M_{f_i}\big)f^n_{iR}(y)=e^{-\sigma}m_f^n\,f^n_{iL}(y)\,,\nonumber\\
&\big(\partial_y+\frac{1}{2}\sigma'+M_{f_i}\big)f^n_{iL}(y)=-e^{-\sigma}m_f^n\,f^n_{iR}(y)\,,
\label{bulk_11}
\end{eqnarray} 
where $\sigma'=\partial_y\sigma$, and a boundary condition 
\begin{eqnarray}
&\left. \delta f^n_{iL}(y)\,f^n_{iR}(y)-\delta f^n_{iR}(y)\,f^n_{iL}(y)\right|^{y=L}_{y=0}=0\,.
\label{bulk_22}
\end{eqnarray} 
Choose $\delta f^n_{iL}(0)=\delta f^n_{iR}(L)=0$,  all the left-handed and the right-handed KK modes of $\Psi_{fi}$ vanish, respectively, at the $y=0$ brane and at the $y=L$ brane. Therefore, only the right-handed (left-handed) modes can couple to the SM fields on the $y=0$ ($y=L$) brane, respectively. The KK fermion equations in Eq.(\ref{bulk_11}), together with the boundary condition, always allow a massless mode. 

Setting $m^f_{n=0}=0$ with $M_{f_i}$ being constant, we get $f^0_{iR}(y)=f^0_{iR}(L)e^{\frac{1}{2}[\sigma(L)-\sigma(y)]-M_{f_i}(L-y)},$
and $f^0_{iL}(y)=f^0_{iL}(0)e^{-\frac{1}{2}\sigma(y)+M_{f_i}y}$.
Rescaling dimensionful operators such as $\hat{Z} \rightarrow e^{\frac{1}{2}\sigma(y)} \hat{Z}$ and $Z \rightarrow e^{-\frac{1}{2}\sigma(y)} Z$, we get  4D Yukawa interactions given as,
\begin{widetext}
\begin{eqnarray}
-{\cal L}_{4D}^q&=&\Big[y^u_{11}\big({\cal U}^c\eta\frac{\Phi_T\sigma}{M^3_5}\big)_{{\bf 1}}+\tilde{y}^u_{11}\big({\cal U}^c\eta\frac{\eta\eta}{M^3_5}\big)_{{\bf 1}}\Big]Q_1H_u+\Big[y^u_{22}({\cal U}^c\frac{\eta}{M_5})_{{\bf 1}''}+\tilde{y}^u_{22}\big({\cal U}^c\frac{\eta\Phi_T\sigma}{M^3_5}\big)_{{\bf 1}''}+\bar{y}^u_{22}\big({\cal U}^c\frac{\eta\eta\eta}{M^3_5}\big)_{{\bf 1}''}\Big]Q_2H_u\nonumber\\
&+&{\bf y}^t_{33}t^cQ_3H_u+{\bf y}^b_{33}b^cQ_3H+\Big[y^d_{22}({\cal D}^c\frac{\eta}{M_5})_{{\bf 1}''}
  + \tilde{y}^d_{22}\big({\cal D}^c\frac{\eta\Phi_T\sigma}{M^3_5}\big)_{{\bf 1}''}+\bar{y}^d_{22}\big({\cal D}^c\frac{\eta\eta\eta}{M^3_5}\big)_{{\bf 1}''}\Big] Q_2 H_d \nonumber \\
&+&\Big[y^d_{11}\big({\cal D}^c\frac{\eta\Phi_T\sigma}{M^3_5}\big)_{\bf 1}+\tilde{y}^d_{11}\big({\cal D}^c\frac{\eta\eta\eta}{M^3_5}\big)_{\bf 1}\Big]Q_1H_d
+{\bf y}^b_{32}b^c(\frac{\Phi_S\Phi_S}{M_5^2})_{{\bf 1}'}Q_2H_d+{\bf y}^b_{31}b^c(\frac{\Phi_S\Phi_S}{M_5^2})_{{\bf 1}''}Q_1H_d\nonumber\\
&+&\Big[y^d_{21}({\cal D}^c\frac{\eta}{M_5})_{{\bf 1}''}+\tilde{y}^d_{21}\big({\cal D}^c\eta\frac{\Phi_T\sigma}{M^3_5}\big)_{{\bf 1}''}
+\bar{y}^d_{21}\big({\cal D}^c\eta\frac{\eta\eta}{M^3_5}\big)_{{\bf 1}''}\Big](\frac{\Phi_S\Phi_S}{M_5^2})_{{\bf1}'}Q_1H_d
\label{4d_a0}\\
-{\cal L}_{4D}^l&=&
\Big[y_e\big(L\frac{\Phi_T\Phi_T}{M^2_5}\big)_{\bf 1}
+\tilde{y}_e\big(L\frac{\Phi_T\Phi_T}{M^2_5}\big)_{\bf 3}\frac{\Phi_T\sigma}{M^2_5}\Big]e^cH_d+\Big[y_\mu(L\Phi_T)_{{\bf 1}'}\frac{\sigma^2}{M^3_5}+\tilde{y}_\mu \Big(L\frac{\eta\eta}{M^3_5}\Big)_{{\bf 1}'}\sigma\Big] \mu^cH_d \nonumber \\
&+& \Big[y_\tau(L\Phi_T)_{{\bf 1}''}\frac{\sigma^2}{M^3_5}+\tilde{y}_\tau\Big(L\frac{\eta\eta}{M^3_5}\Big)_{{\bf 1}''}\sigma \Big]\tau^c H_d 
+y_\nu(N^cL)_{\bf 1}\frac{\Theta}{M_5}H_u+\tilde{y}_\nu(N^cL)_{\bf 3}\frac{\Phi_S}{M_5}H_u \nonumber \\
&+&\frac{1}{2}(y_\Theta\Theta+y_{\tilde{\Theta}}\tilde{\Theta})(N^cN^c)_{\bf 1}+\frac{1}{2}y_R\Phi_S (N^cN^c)_{\bf 3}+{\rm h.c.}
\label{4d_a}
\end{eqnarray}
\end{widetext}
with well-defined Yukawa-coupling functions
%\begin{widetext}
\begin{eqnarray}
& {\bf y}^\alpha_{ij}=4Z^\omega_{ij}\hat{Z}_{\alpha k}\,e^{-M_{\omega_k}L}\cosh\sigma(L)/2\,,\nonumber\\
&y^\alpha_{ij}/M_5=4Z^\omega_{ij}Z_{\alpha k}\,e^{-M_{\omega_k}L}\cosh\sigma(L)/2\,,\nonumber\\
&y_{\ell}/M_5=4Z_{\ell} Z_L\,e^{-M_\ell L}\cosh\sigma(L)/2\,,
\label{np_1}
\end{eqnarray}
%\end{widetext}
where $\alpha=(d,s,b,u,c,t)$, $\omega=(u, d)$, and $\ell=(e, \mu, \tau)$.%, and the Yukawa couplings of capital letter $Z$ are functions of flavons, ${\cal F}/M_5$.

\section{Hierarchies of fermion masses and mixings}
\subsection{Charged fermions}
After electroweak symmetry breaking, we get the VEVs for two Higgs doublets as $\langle H_{u,d}\rangle=v_{u,d}$ where $v_{u}=v\sin\beta/\sqrt{2}$ and $v_{d}=v\cos\beta/\sqrt{2}$ with $v=246$ GeV.
Given the specific vacuum alignment given by  Eqs.\,(\ref{vev1},\ref{vev2}), we obtain the up(down)-type quark mass matrices and diagonal form of charged-lepton mass matrix, in which each entries is proportional to $e^{-M_{f_i}L}\cosh\sigma(L)/2$, see Eq.(\ref{ChL1}). 

From the diagonalization of the quark mass matrices, we finally obtain the quark mixings $\theta^q_{ij}$ in the standard form\cite{PDG} corresponding to the rotations in the $ij=12, 23, 13$ planes, as follows;
\begin{eqnarray}
\theta^q_{ij}&\simeq& 3 \kappa^2 f(\hat{\bf y},\hat{y}) \Big|\frac{\delta^{\rm G}_1}{\delta^{\rm G}_2}\Big(\frac{2}{1+\kappa^2}\Big)^{\frac{1}{2}}\nabla_\chi\Big|^{[{\cal Q}_{X_1}(Q_j)-{\cal Q}_{X_1}(Q_i)]}\,,\nonumber\\
\delta^{q}_{\scriptscriptstyle{CP}}&=&\arg(f(\hat{\bf y},\hat{y}))\,,
\label{qckm}
 \end{eqnarray}
where $i\neq j=1,2,3$ and ${\cal Q}_{X_i}(q)$ represents $U(1)_{X_i}$ quantum number ${\cal Q}$ of field $q$, and $f_i(\hat{\bf y},\hat{y})$ are functions of the associated {\it hat} Yukawa couplings being of order unities (cf. Eqs.(\ref{4dL},\ref{np_1}, \ref{ChL1})). 
Note that in the limit of $\kappa \rightarrow 0$ there is no quark mixing.
Clearly, it shows the quark mixings only depend on the local effects of $G_F$. 
For the charged lepton mass ratios are given by 
 \begin{eqnarray}
&\frac{m_e}{m_\mu}\simeq\hat{y}_{e}\nabla^2_T\,\nabla^{10}_\chi/(\hat{y}_{\mu}\nabla_T\nabla_\varrho+\hat{\tilde{y}}_{\mu}\nabla^2_\eta)\nabla_\varrho\,,\nonumber\\
&\frac{m_\mu}{m_\tau}=(\hat{y}_{\mu}\nabla_T\nabla_\varrho+\hat{\tilde{y}}_{\mu}\nabla^2_\eta)\nabla^{4}_\chi/(\hat{y}_{\tau}\nabla_T\nabla_\varrho+\hat{\tilde{y}}_{\tau}\nabla^2_\eta)\,.
\label{lm}
\end{eqnarray}
In the limit, $ \nabla_T\sim \nabla_\varrho \sim \nabla_\eta$, and for order one $\hat{y}_{\alpha=e,\mu,\tau}$ and $\hat{\tilde{y}}_{\alpha}$,
$\frac{m_e}{m_\mu}\simeq \nabla^{10}_\chi \nabla_{\varrho}$ and $\frac{m_\mu}{m_\tau}\simeq \nabla^{4}_\chi$.

\subsection{Neutrinos}
In this model, the low energy effective neutrino masses are generated by the usual seesaw mechanism\cite{Minkowski:1977sc} with the inclusion of right-handed $SU(2)$-singlet Majorana neutrinos $N^c$.
The seesaw scale $M$ should be larger than the electroweak scale but smaller than $M_5$.
The fundamental gravity scale $M_5$ can be derived in terms of $F_A$ or $M$:
{\small\begin{eqnarray}
M_5=\frac{|\delta^{\rm G}_2|}{\sqrt{2}}\frac{F_A}{\nabla_\chi}=\frac{M}{\nabla_\Theta}\quad\text{with} ~M\equiv|\delta^{\rm G}_1|\frac{F_A}{\sqrt{1+\kappa^2}}\,,
\label{gr5}
\end{eqnarray}}
where $F_A$ is the QCD axion decay constant, given by $F_A\equiv\frac{f_A}{N_C}=\frac{f_{a_i}}{\sqrt{2}\,\delta^{\rm G}_i}$ with $f_{a_1}=\sqrt{2}v_\chi$ and $f_{a_2}=v_\Theta\sqrt{1+\kappa^2}$.
Contrary to the quark and charged-lepton sectors, the neutrino Yukawa couplings do not depend on the non-local effects of the extra-dimension.
After seesawing ${\cal M}_{\nu}\simeq -m^T_DM^{-1}_Rm_D$ where $M_R$ and $m_D$ are respectively right-handed Majorana and Dirac neutrino mass matrices, in a basis where charged lepton masses are real and diagonal, we obtain
the effective light neutrino mass matrix
\begin{widetext}
 \begin{eqnarray}
{\cal M}_{\nu}&=&m_0\frac{e^{i\pi}}{3}{\left(\begin{array}{ccc}
  1 & 1 & 1  \\
  1 & 1 & 1 \\
 1 & 1 & 1
 \end{array}\right)}+\frac{m_0\,e^{i\pi}}{1-e^{2i\phi}\tilde{\kappa}^2}{\left(\begin{array}{ccc}
  a_\nu &  -\frac{1}{2}a_\nu &  -\frac{1}{2}a_\nu \\
   -\frac{1}{2}a_\nu & a_\nu-1 &1-\frac{1}{2}a_\nu \\
 -\frac{1}{2}a_\nu &1-\frac{1}{2}a_\nu  & a_\nu-1
 \end{array}\right)} \nonumber\\
 &+&m_0\,e^{i\pi}\frac{\kappa^2}{1-e^{2i\phi}\tilde{\kappa}^2}{\left(\begin{array}{ccc}
  - b_\nu - c_\nu\,\tilde{\kappa} &  \frac{1}{2}(b_\nu+c_\nu\tilde{\kappa})+ d_\nu &  \frac{1}{2}(b_\nu+c_\nu\tilde{\kappa})- d_\nu \\
   \frac{1}{2}(b_\nu+c_\nu\tilde{\kappa})+ d_\nu & \frac{1}{2}(b_\nu-c_\nu\tilde{\kappa})-d_\nu & \frac{1}{2}c_\nu\tilde{\kappa}-b_\nu \\
 \frac{1}{2}(b_\nu+c_\nu\tilde{\kappa})- d_\nu & \frac{1}{2}c_\nu\tilde{\kappa}-b_\nu  & \frac{1}{2}(b_\nu-c_\nu\tilde{\kappa})+d_\nu
 \end{array}\right)},
 \label{nu_mass}
 \end{eqnarray}
 \end{widetext}
where $m_{0}=|\hat{y}_\nu|^2\frac{v^2_u}{M}\left|\frac{X_2\delta^{\rm G}_1}{X_1\delta^{\rm G}_2}\nabla_\chi\right|^{16}\left(\frac{2}{1+\kappa^2}\right)^8$, $\tilde{\kappa}=\kappa|\hat{y}_{R}/\hat{y}_\Theta|$ with $\phi=\arg(\hat{y}_{R}/\hat{y}_{\Theta})$, and each component is given by $a_\nu$, $b_\nu$, $c_\nu$, $d_\nu$ that are functions of $\kappa$, $\tilde{\kappa}$, $\phi$, and other phases.

As a result, the neutrino masses $m_{\nu_i}$ ($i=1,2,3$) are obtained by the transformation, $U^T_{\rm PMNS}\,{\cal M}_{\nu}\,U_{\rm PMNS}={\rm diag}(m_{\nu_1}, m_{\nu_2}, m_{\nu_3})$, where $U_{\rm PMNS}$ is the mixing matrix of three mixing angles, $\theta_{12}$ (solar), $\theta_{13}$ (reactor), $\theta_{23}$ (atmospheric), and three CP-odd phases (one $\delta_{CP}$ for the Dirac neutrinos and two $\varphi_{1,2}$ for the Majorana neutrinos)\cite{PDG}. In the limit $\kappa\rightarrow0$ (leading to $\tilde{\kappa}\rightarrow0$ and $a_\nu\rightarrow\frac{2}{3}$), the light neutrino masses generated via Eq.(\ref{nu_mass}) become degenerate with no neutrino mixings (and also no quark mixings, see Eq.(\ref{qckm})).
Hence, it is reasonable to take a nonzero $\kappa$ for generating the observed neutrino mixing angles. 
Then, the neutrino mass eigenvalues of Eq.(\ref{nu_mass}) can be expanded in terms of $\kappa$:  for normal mass ordering (NO), $m^2_{\nu_3}=m^2_{\nu+}>m^2_{\nu_2}\equiv m^2_0>m^2_{\nu_1}=m^2_{\nu-}$ with
%\begin{eqnarray}
  $m^2_{\nu\pm}=m^2_0\big(Q_\nu\pm\sqrt{J_\nu}\big)/P_\nu$,
 %\label{numa}
 %\end{eqnarray}
and for inverted mass ordering (IO), $m^2_{\nu_2}\equiv m^2_0>m^2_{\nu_1}=m^2_{\nu+}>m^2_{\nu_3}=m^2_{\nu-}$,
where $Q_\nu$, $J_\nu$, and $P_\nu$ are functions of the parameters in Eq.(\ref{nu_mass}).
The limit $\tilde{y}_\nu\rightarrow0$ in Eq.(\ref{4d_a}) is equivalent to $a_\nu\rightarrow\frac{2}{3}(1-e^{i\phi}\tilde{\kappa})$ and $\{b_\nu, c_\nu, d_\nu\}\rightarrow0$, and it gives rise to the so-called tri-bimaximal mixing of neutrinos.
Then, the deviation from the tri-bimaximal mixing\cite{TBM} can be presented in terms of $\kappa$:
  \begin{eqnarray}
 &|\theta_{23}-\pi/4|\propto|d_\nu|\kappa^2\,,\qquad\theta_{13}\propto|d_\nu|\kappa^2\,,\nonumber\\
 & \theta_{12}-\frac{1}{2}\tan^{-1}(2\sqrt{2})\simeq\frac{2\sqrt{2}|3|a_\nu|(\kappa^2-\tilde{\kappa}^2)\,S_\nu+{\cal O}(\kappa^3,\tilde{\kappa}^4)|}{R_\nu+3|a_\nu|\big(\tilde{\kappa}^2-\frac{\kappa^2}{2}\big)\,S_\nu+{\cal O}(\kappa^3,\tilde{\kappa}^4)}\,,
 \end{eqnarray}
 where $R_\nu=\frac{1}{3}-\frac{3}{4}|a_\nu|^2-\frac{2}{3}\tilde{\kappa}^2\cos^2\phi$, and $S_\nu$ is functions of $|b_\nu|, |c_\nu|, \tilde{\kappa}$ and relevant phases (especially, $S_\nu\rightarrow0$ and ${\cal O}(\kappa^3,\tilde{\kappa}^4)\rightarrow0$ for $\{b_\nu,c_\nu\}\rightarrow0$).

%%%%%%%%%%%%%%%%%%%%%%%%%%%%%%%%%
\begin{figure}[h]
\begin{center}
\includegraphics*[width=0.35\textwidth]{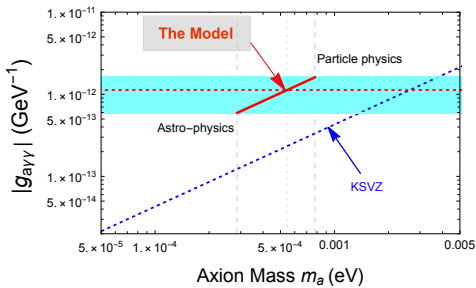}
\caption{\label{Fig1} Plot of $|g_{a\gamma\gamma}|$ versus $m_{a}$ for KSVZ\cite{KSVZ} (blue-dotted line) and the model (``localized" red line) in terms of $E_A/N_C=$ $0$ and $-761/99$, respectively.}
\end{center}
\end{figure}
%%%%%%%%%%%%%%%%%%%%%%%%%%

\begin{figure}[h]
%\vspace*{-5.0cm}
%\hspace*{-1cm}
\begin{minipage}[h]{9.3cm}
\epsfig{figure=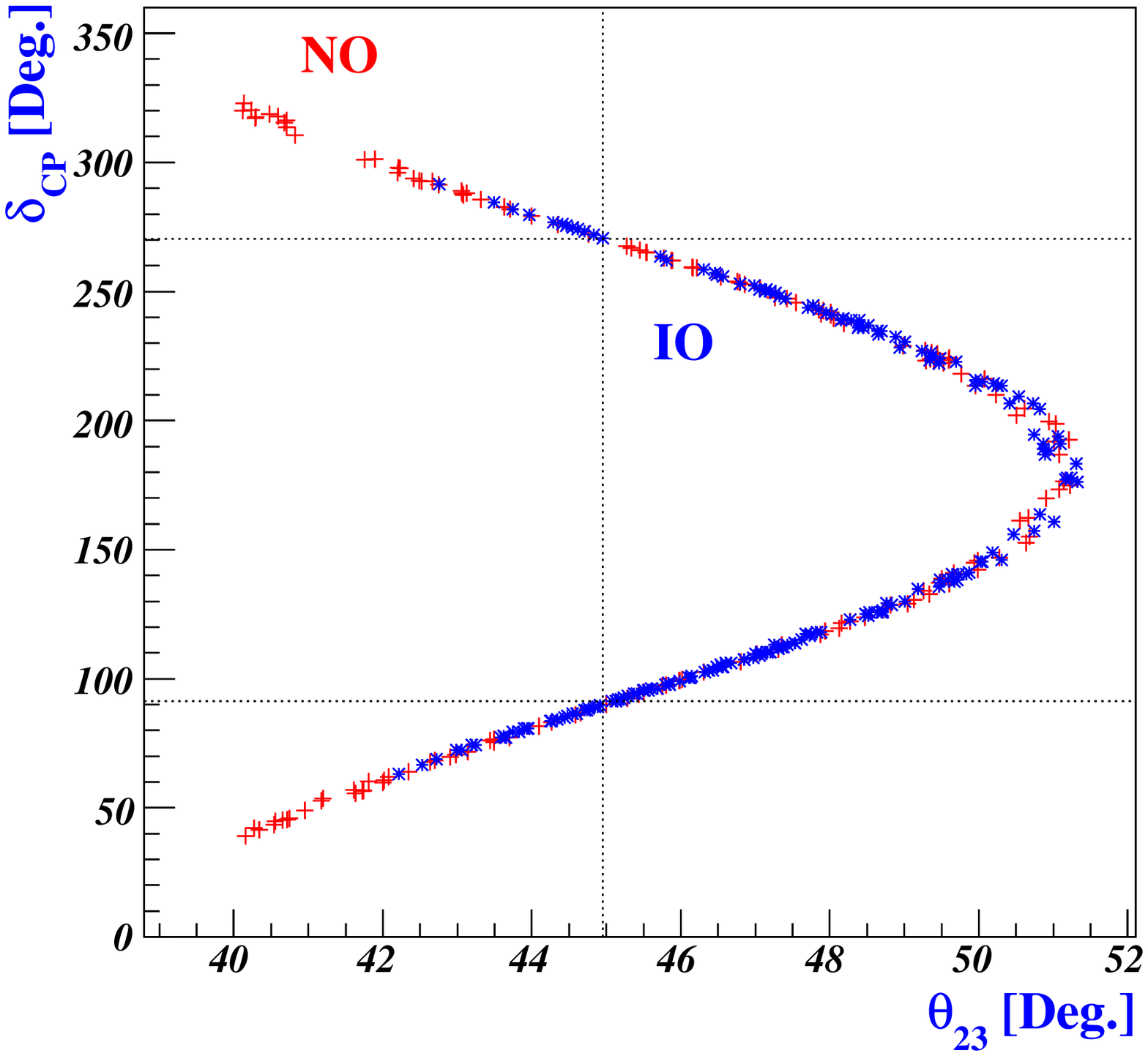, width=4.4cm,angle=0}
%\end{minipage}
%\hspace*{1.0cm}
%\begin{minipage}[h]{7.3cm}
\epsfig{figure=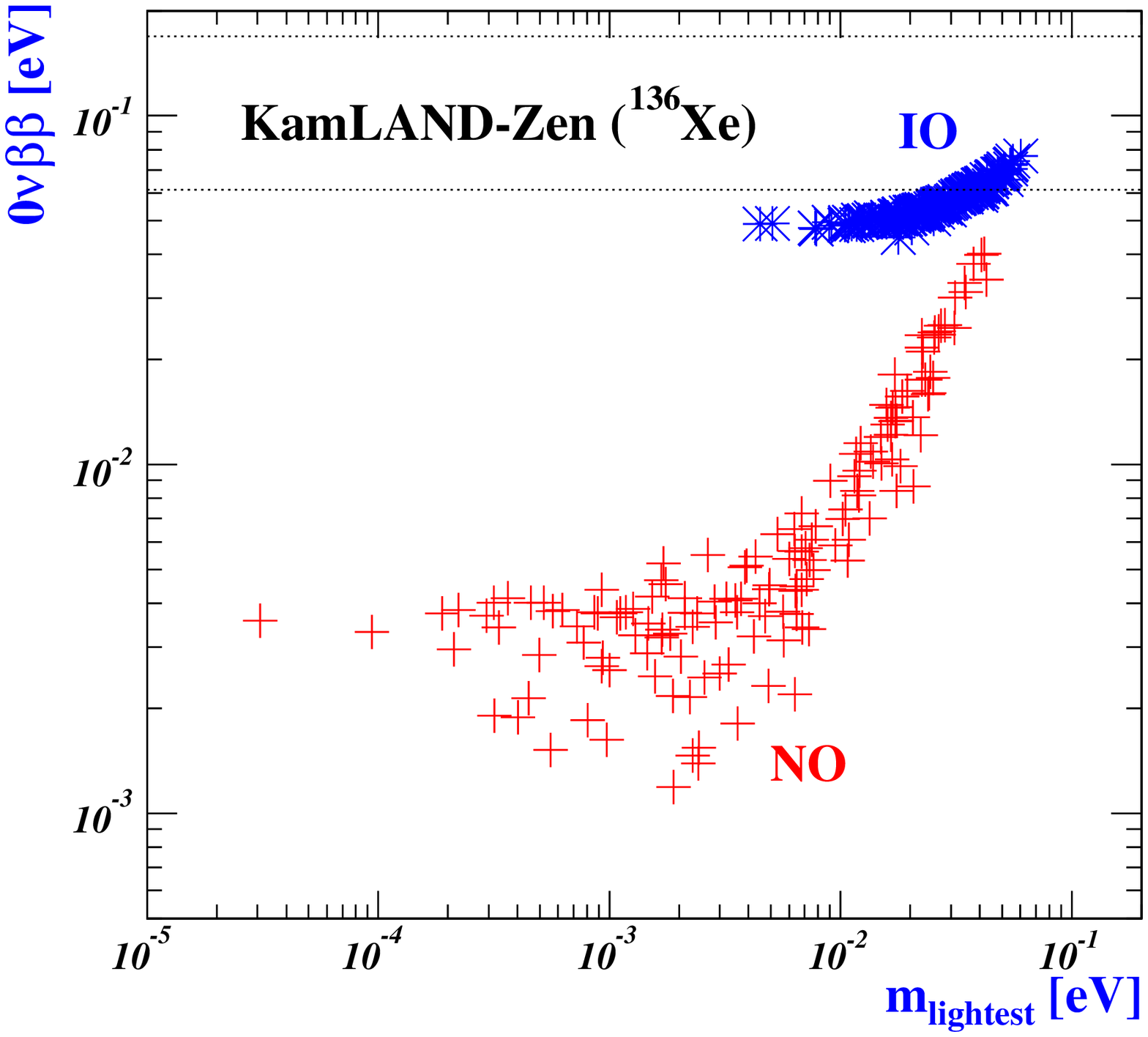,width=4.4cm,angle=0}
\end{minipage}
%\vspace*{-5.5cm}
\caption{\label{Fig2} Left-plot for predictions of $\delta_{CP}$ as a function of $\theta_{23}$, for NO (red crosses) and IO (blue asters). Right-plot for $0\nu\beta\beta$-decay rate as a function of the lightest neutrino mass, $m_{\rm lightest}$, for NO ($m_{\rm lightest}=m_1$ red crosses) and IO ($m_{\rm lightest}=m_3$ blue asters), where the most stringent limit ($90\%$ CL) is given by KamLAND-Zen\cite{KamLAND-Zen:2016pfg}.}
\end{figure}

\section{Numerical results}
Using the quark and charged-lepton masses, we can determine the warp factor $\sigma(L)=kL$  with the help of Eq.(\ref{re_M}) via flavor dynamics. In the end, the compactification length, $L[m]$, is given by $L[m]\simeq5.82\times10^{-13}\,\cdot\frac{e^{-\sigma(L)}\sigma(L)}{\sinh\sigma(L)}\big(\frac{10^{11}{\rm GeV}}{M_5}\big)^3$. For $\hat{y}(\hat{\bf y})={\cal O}(1)$, there are fourteen degrees of freedom left due to the two constraints in Eq.(\ref{ab}) among 16 parameters (8 local parameters $\{\nabla_\chi,\nabla_\Theta,\nabla_S, \kappa, \nabla_\eta,\nabla_T, \nabla_{\varrho}, \tan\beta\}$  plus 8 non-local parameters $\{\sigma(L), M^{u(d)}_{1,2,3}, M_\ell\}$).  Thirteen observables in the charged-fermion sectors can be determined once one of non-local parameters is fixed.

We perform a numerical simulation by using the linear algebra tools\cite{Antusch:2005gp}. With inputs $\sigma(L)=18.25$, $\tan\beta=3.1$, $\kappa=0.4$, $\nabla_\chi=0.52$, $\nabla_\eta=0.51$, $\nabla_T=0.35$, $\nabla_\varrho=0.34$, the empirical results of the SM fermion mixing angles and masses\cite{PDG} are obtained with the appropriate  Yukawa couplings $\hat{y}$ of order of unity  of Eq.(\ref{4dL}). Plugging the inputs into Eq.(\ref{gr5}), we can derive the scale of $M_5$ as $(1-2)\times10^{11}\,{\rm GeV}\lesssim M_{5}\lesssim2.8\times10^{11}\,{\rm GeV}$, leading to the QCD axion mass, $m_{a}=5.44^{+2.33}_{-2.58}\times10^{-4}\,\text{eV}$. Then, we obtain the axion photon coupling $|g_{a\gamma\gamma}|=1.13^{+0.48}_{-0.53}\times10^{-12}\,\text{GeV}^{-1}$ for $m_u/m_d=0.47$, as shown in red line in FIG.\,\ref{Fig1}. Moreover, we can predict $\delta_{CP}$ precisely by $\theta_{23}$. Namely, $\theta_{23}$ would favor $\sim51^{\circ}$ for $\delta_{CP}\sim180^{\circ}$ and $\theta_{23}\sim45^{\circ}$ for $\delta_{CP}\sim90^{\circ}$ and $270^{\circ}$, as shown in the left plot of FIG.\,\ref{Fig2}. 
We note that the pattern of the light neutrino mass spectrum, NO ($m_{\nu_3}>m_{\nu_2}>m_{\nu_1}$) or IO ($m_{\nu_2}>m_{\nu_1}>m_{\nu_3}$), can be distinguished by the measurement of $0\nu\beta\beta$-decay rate, as can be seen in the right plot of FIG.\,\ref{Fig2}.

\section{Flavored axion}
The flavored-axion $A_1$ produces the flavor-changing neutral Yukawa interaction, $-{\cal L}^{aq}\supset i\frac{A_1}{f_{a1}}\lambda(1-\frac{\lambda^2}{2})(m_d-m_s)\bar{d}s+{\rm h.c.}$ where $\lambda$ is the Cabbibo angle\cite{PDG}, which gives the strongest bound on the QCD axion decay constant $F_{A}$. 
From the present experimental upper bound, ${\rm Br}(K^+\rightarrow\pi^+A_i)<(3-6)\times10^{-11}$ at $90\%$ CL\cite{NA62:2021zjw}, we obtain $F_{A}\gtrsim(0.7-1.5)\times10^{10}\,{\rm GeV}$.

Under the $U(1)_{\rm PQ}$ transformation, the  QCD axion field $A$ changes $A\rightarrow A+\frac{f_{A}}{N_C}\alpha$, where $f_A=\sqrt{2}\,\delta^{\rm G}_2f_{a_1}=\sqrt{2}\,\delta^{\rm G}_1f_{a_2}$. From $-{\cal L}^{a\ell}\supset (\frac{\partial_\mu A_1}{2f_{a_1}}-41\frac{\partial_\mu A_2}{2f_{a_2}}+\frac{\partial_\mu A_T}{v_t})\bar{e}\gamma^\mu\gamma_5 e$, we can identify the effective axion couplings as $g_{1ee}=m_e/(\sqrt{2}|\delta^{\rm G}_1|F_A)$ and $g_{2ee}=41m_e/(\sqrt{2}|\delta^{\rm G}_2|F_A)$.
Thus, the flavored-axions $A_{1,2,T}$ interactions with leptons can be searched for in stellar evolutions in astroparticle physics\cite{Raffelt:1985nj}. 
Some stellar cooling hints can be interpreted as axion-electron couplings $7.2\times10^{-14}\lesssim|g_{iee}|\lesssim2.2\times10^{-13}$\cite{Rowell:2011wp}, determining the favored values of $F_A$ via the $A_i$ couplings to electrons.
 Otherwise, the lower limit on $F_A$ is set to about $10^8\,{\rm GeV}$ by SN1987A while the upper bound on $F_A$ is about $10^{12}\,{\rm GeV}$ from the dark matter abundance. 
Combining the stellar cooling hints (from astro-physics) with the constraint from $K^+\rightarrow\pi^++A_1$ process (from particle physics), we obtain the consistent axion decay constant as $(0.7-1.5)\times10^{10}\lesssim F_A[{\rm GeV}]\lesssim1.9\times10^{10}$.

\section{Conclusion}
 We have proposed an extra-dimension scenario for understanding the origin of the fermion mass and mixing hierarchies by
introducing the localized flavon fields and imposing the flavor symmetry $G_F(=${\it non-Abelian}$\times${\it Abelian}) through the bulk.
We fixed the charges of the extra gauged $U(1)$ symmetries by the U(1) gravitation anomaly-free condition and found that they play a crucial role in achieving the desirable fermion mass and mixing hierarchies and protecting the PQ symmetry from quantum gravity corrections.

When the bulk flavor symmetry $G_F$ is broken due to the flavor fields localized at the 3-branes, we showed that the $SU(2)$ singlet flavored fermions  in the bulk are integrated out to provide the effective Yukawa couplings for quarks and leptons, inherited with the $G_F$ breaking in the two sectors. We also showed that there is a viable parameter space for the flavored axion in our model, which is consistent with the solution to the strong CP problem, the dark matter relic abundance with misalignment mechanism as well as the bounds from rare meson decays.

%\newpage
%\onecolumngrid

\appendix

%%%%%%%%%%%%%%%%%%%%%%%%%%%%%%%%%%%%%%%%%%%%%%%%%%%%%%%%%%%%%%%%%%%%%%%%%%%%%%%%%%%%%%%%%%%%%%%%%%
\section{Superpotential}
\label{sec:appen1}
We present the vacuum structure for the flavon fields in our model.

In order to obtain the relevant vacuum configuration for the desirable lepton and quark mixings, several SM singlet fields are introduced in Table-\ref{reps_v}: flavon fields ${\cal F}$ has a zero $U(1)_R$ charge and transforms nontrivially only under $G_F$ and are responsible for the symmetry breaking of $G_F$; driving fields has $U(1)_R$ charge $+2$ and breaks the flavor group along the required VEV directions.
 The vacuum configuration for flavons is simplified in a supersymmetric theory.
 Thus, we consider the brane-localized superpotential with the driving fields at the leading order, which is invariant under $SL_2(F_3)\times U(1)_{X_i}$ ($i=1,2, T$), as follows,
% \begin{widetext}
\begin{eqnarray}
&W_v=\delta(y)\big\{\Phi^S_0(g_{s_1}\Phi_S\Phi_S+g_{s_2}\tilde{\Theta}\Phi_S)\nonumber\\
&+\Theta_0(g_{\Theta_1}\Phi_S\Phi_S+g_{\Theta_2}\Theta\Theta+g_{\Theta_3}\Theta\tilde{\Theta}+g_{\Theta_4}\tilde{\Theta}\tilde{\Theta})\big\}\nonumber\\
&+\delta(y-L)\big\{\Phi^T_0\big(\mu_T\Phi_T+g_T\Phi_T\Phi_T\frac{\sigma}{M_5}+g_{T\chi}\Phi_T\frac{\chi\tilde{\chi}}{M_5}\big)\nonumber\\
&+\sigma_0\big(\mu_\sigma\sigma+g_{\sigma\chi}\chi\tilde{\chi}\frac{\sigma}{M_5}\big)\nonumber\\
&+ \eta_0\big(\mu_\eta\eta+g_\eta\eta\Phi_T\frac{\sigma}{M_5}\big)+\chi_0\big(g_{\chi}\chi\tilde{\chi}-\mu^2_\chi\big)\big\},
\label{d_pot}
\end{eqnarray}
%\end{widetext}
where $\{\Phi_0, \Theta_0, \eta_0,\sigma_0, \chi_0\}$ denote the driving fields, and $\mu_{i}$ are dimensionful parameters and $g_{i}$ are dimensionless coupling constants. At this level there is no fundamental distinction between the $\Theta$ and $\tilde{\Theta}$ in $W_v$ so that the field $\tilde{\Theta}$ is defined as the combination that couples to $\Phi^S_0\Phi_S$\cite{Altarelli:2005yx}.  Higher-dimensional brane interactions via the one-loop exchange of the flavored bulk fermions are allowed but absorbed into the leading order terms of Eq.(\ref{d_pot}) by redefinition of coefficients.

\section{Fermion mass matrices}
We present the details of quark and lepton mass matrices obtained in 4D effective theory.

From Eqs.(\ref{4d_a0}) and (\ref{4d_a}) the up(down)-type quark ${\cal M}_{u(d)}$ and charged-lepton ${\cal M}_{\ell}$ mass matrices are expressed as
\begin{widetext}
 \begin{eqnarray}
 &{\cal M}_{d}={\left(\begin{array}{ccc}
 m^d_{11}\,e^{i(3\frac{A_1}{v_{\cal F}}+14\frac{A_{2}}{v_{g}})}\,e^{-M_{d_1}L} &  0 &  0 \\
 m^d_{21}\,e^{i(3\frac{A_1}{v_{\cal F}}+14\frac{A_{2}}{v_{g}})}\,e^{-M_{d_2}L}   &  m^d_{22}\,e^{i(\frac{A_1}{v_{\cal F}}+14\frac{A_{2}}{v_{g}})}\,e^{-M_{d_2}L}  &  0   \\
 m^d_{31}\,e^{i(11\frac{A_1}{v_{\cal F}}-18\frac{A_{2}}{v_{g}})}\,e^{-M_{d_3}L}   &  m^d_{32}\,e^{i(9\frac{A_1}{v_{\cal F}}-18\frac{A_{2}}{v_{g}})}  & m^d_{33}\,e^{i(3\frac{A_1}{v_{\cal F}}-18\frac{A_{2}}{v_{g}})}\,e^{-M_{d_3}L} 
 \end{array}\right)}\cosh\frac{\sigma(L)}{2}\,,  \nonumber\\
& {\cal M}_{u}={\rm diag}\big(
m^u_{11}\,e^{i(2\frac{A_1}{v_{\cal F}}+6\frac{A_{2}}{v_{g}})}\,e^{-M_{u_1}L},  m^u_{22}e^{6i\frac{A_2}{v_g}}\,e^{-M_{u_2}L}, m^u_{33}\,e^{-11i\frac{A_2}{v_g}}\,e^{-M_{u_3}L}\big)\cosh\frac{\sigma(L)}{2}, \nonumber\\
&{\cal M}_{\ell}={\rm diag}\big(m^e_{11}\,e^{i(\frac{A_1}{v_{\cal F}}-41\frac{A_{2}}{v_{g}}+2\frac{A_T}{v_t})}, m^{\mu}_{22}\,e^{i(\frac{A_1}{v_{\cal F}}-31\frac{A_{2}}{v_{g}}-\frac{A_T}{v_t})} , m^{\tau}_{33}\,e^{i(\frac{A_1}{v_{\cal F}}-27\frac{A_{2}}{v_{g}}-\frac{A_T}{v_t})}\big)e^{-M_\ell L}\cosh\frac{\sigma(L)}{2},
 \label{ChL1}
  \end{eqnarray}
\end{widetext}
with the massless modes $A_{1(2, T)}$ and $v_t=\sqrt{v^2_T+v^2_\varrho}$, where 
 \begin{eqnarray}
&m^d_{11}= (i\hat{y}^d_{11}\nabla_T\nabla_\varrho-\hat{\tilde{y}}^d_{11}\nabla^2_\eta)\nabla_\eta|\frac{\delta^{\rm G}_1}{\delta^{\rm G}_2}|^3(\frac{2}{1+\kappa^2})^{\frac{3}{2}}\,\nabla^{17}_{\chi}\,v_d, \nonumber\\
&m^d_{21}\simeq-3\kappa^2\hat{y}^d_{21}\nabla_\eta|\frac{\delta^{\rm G}_1}{\delta^{\rm G}_2}|^3(\frac{2}{1+\kappa^2})^{\frac{3}{2}}\,\nabla^{17}_{\chi}\,v_d,\nonumber\\ 
&m^d_{22}\simeq -\hat{y}^d_{22}\nabla_\eta|\frac{\delta^{\rm G}_1}{\delta^{\rm G}_2}|(\frac{2}{1+\kappa^2})^{\frac{1}{2}}\,\nabla^{15}_{\chi}\,v_d\nonumber\\
 &m^d_{31}=3\kappa^2 \hat{\bf y}^b_{31}|\frac{\delta^{\rm G}_1}{\delta^{\rm G}_2}|^{11}(\frac{2}{1+\kappa^2})^{\frac{11}{2}}\,\nabla^{29}_{\chi}\,v_d,\nonumber\\
 &m^d_{32}=3\kappa^2 \hat{\bf y}^b_{32}|\frac{\delta^{\rm G}_1}{\delta^{\rm G}_2}|^9(\frac{2}{1+\kappa^2})^{\frac{9}{2}}\,\nabla^{27}_{\chi}\,v_d, \nonumber\\ 
 &m^d_{33}= \hat{\bf y}^b_{33}|\frac{\delta^{\rm G}_1}{\delta^{\rm G}_2}|^3(\frac{2}{1+\kappa^2})^{\frac{3}{2}}\,\nabla^{21}_{\chi}\,v_d,
    \end{eqnarray}
 \begin{eqnarray}
&m^u_{11}=(i\hat{y}^u_{11}\nabla_T\nabla_\varrho-\hat{\tilde{y}}^u_{11}\nabla^2_\eta)\nabla_\eta|\frac{\delta^{\rm G}_1}{\delta^{\rm G}_2}|^2\frac{2}{1+\kappa^2}\,\nabla^8_{\chi}\,v_u,\nonumber\\
&m^u_{22}\simeq-\hat{y}^u_{22}\nabla_\eta\,\nabla^6_{\chi}\,v_u, \nonumber\\ 
&m^{u}_{33}=\hat{\bf y}^{t}_{33}\,\nabla^{11}_{\chi}\,v_u,
 \end{eqnarray}
  \begin{eqnarray}
 &m^e_{11}\simeq\hat{y}_{e}\nabla^2_T\,\nabla^{42}_\chi|\frac{\delta^{\rm G}_1}{\delta^{\rm G}_2}|(\frac{2}{1+\kappa^2})^{\frac{1}{2}}\,v_{d}, \nonumber\\ 
 &m^\mu_{22}=(\hat{y}_{\mu}\nabla_T\nabla_\varrho+\hat{\tilde{y}}_{\mu}\nabla^2_\eta)\nabla^{32}_\chi\nabla_\varrho|\frac{\delta^{\rm G}_1}{\delta^{\rm G}_2}|(\frac{2}{1+\kappa^2})^{\frac{1}{2}}\,v_{d},\nonumber\\
 &m^\tau_{33}=(\hat{y}_{\tau}\nabla_T\nabla_\varrho+\hat{\tilde{y}}_{\tau}\nabla^2_\eta)\nabla_\varrho \nabla^{28}_\chi|\frac{\delta^{\rm G}_1}{\delta^{\rm G}_2}|(\frac{2}{1+\kappa^2})^{\frac{1}{2}} \,v_{d}.
 \end{eqnarray}
 
%%%%%%%%%%%%%%%%%%%%%%%%%%%%%%%%%%%%%%%%%%%%%%%%%%%%%%%%%%%%%%%%%%%%%%%%%%%%%%%%%%%%%%%%%%%%%%%%%%
\acknowledgments
{YHA was supported by the National Research Foundation of Korea(NRF) grant funded by the Korea government(MSIT) (No.2020R1A2C1010617).
SKK was supported by the National Research Foundation of Korea(NRF) grant funded by the Korea government(MSIT) (No.2019R1A2C1088953).
 HML was supported in part by Basic Science Research Program through the National Research Foundation of Korea (NRF) funded by the Ministry of Education, Science and Technology (NRF-2019R1A2C2003738 and NRF-2021R1A4A2001897). 
}

%%%%%%%%%%%%%%%%%%%%%%%%%%%%%%%%%%%%%%%%%%%%%%%%%%%%%%%%%%%%%%%%%%%%%%%%%%%%%%%%%%%%%

\end{document}